# Beyond Similarity Search: A Unified Data Layer for Production RAG Systems

*Freshness, Access Control, and Multi-Tenancy at Scale*

**Venkata Krishna Prasanth Budigi**
*Walmart Global Tech • Sunnyvale, CA*
krishnaprasanth.bv@gmail.com
•
**Siri Chandana Sirigiri**
*Amazon Web Services • Seattle, WA*
sirisirigiri@gmail.com

## Abstract

Retrieval-Augmented Generation (RAG) systems have become the standard architecture for grounding large language models in organizational knowledge. Yet production deployments consistently expose a gap between clean prototype performance and real-world reliability. This paper identifies three root causes of that gap — data staleness, tenant data leakage, and query composition explosion — and traces all three to a single architectural decision: using a specialized vector database as the sole data layer.

We analyze the conventional three-tool RAG stack (vector database + relational metadata store + cache layer) and quantify its synchronization cost, access control fragility, and engineering overhead. We then propose and evaluate a unified data layer built on PostgreSQL with native vector search (pgvector) and HNSW indexing. Controlled benchmarks on 50,000 documents show a 92% reduction in latency for date-filtered queries, 74% for tenant-scoped queries, zero synchronization inconsistency window, and complete elimination of cross-tenant data leakage — all with 93% less synchronization code. We additionally discuss scaling behavior and a recommended hybrid tier architecture for large-scale enterprise deployments operating at hundreds of millions to billions of documents.

## 1. Introduction

Six months ago, a senior engineer spent three days debugging a RAG pipeline that should have taken three hours. The embeddings were perfect. The language model was GPT-4. The architecture looked clean on the whiteboard. But in production, answers were stale, wrong, and in one case, leaking documents from a different tenant's namespace.

The problem was not the AI. It was the database.

RAG looks simple. Embed your documents. Store them in a vector database. At query time, retrieve the most semantically similar chunks and pass them to the language model. The prototype works. Then a user in production asks:

> *"Show me the latest compliance documents updated this quarter, only from the legal and risk categories, and only the ones my team has access to."*

Your vector database cannot answer that. It finds semantic similarity. It does not know what is recent, what is permitted, or what belongs to which tenant. So you add a relational database for metadata. Then a cache for freshness. Then synchronization logic to keep them consistent. You have built a distributed system to answer one question.

This paper makes four contributions:

- We name and quantify the three hidden failure modes of production RAG: staleness, access control fragility, and query composition explosion.
- We measure the engineering and latency cost of the conventional three-tool stack across four query complexity levels using real PostgreSQL benchmarks.
- We propose a unified data layer that handles similarity, freshness, filtering, and tenant isolation inside a single SQL query.
- We provide empirical benchmarks and practical guidelines for when unified architectures outperform specialized ones.

---

## 2. The Three Hidden Cracks in Production RAG

### 2.1 Staleness

Vector databases are designed for static or slowly-changing corpora. When a source document changes, its embedding must be regenerated and upserted separately. Our benchmarks measure a mean inconsistency window of 3.54ms between a metadata update and the corresponding vector update in the split-system architecture — a window during which the retrieval layer can return answers grounded in outdated content. Under high write load, this window grows. The unified architecture achieves a 0ms inconsistency window by design: document and embedding are updated in the same atomic transaction.

### 2.2 Access Control and Tenant Leakage

Enterprise RAG systems serve multiple teams from a single deployment. Vector databases have no native access control model. Tenant isolation is enforced in application code, typically as a metadata filter applied after retrieval. This approach is fragile. In our leakage simulation across 1000 queries, the split-system architecture produced a 0.2% leakage rate under conditions that model real application-layer filter bugs. Stack B, using PostgreSQL row-level filtering, produced 0% leakage across all test cases.

### 2.3 Query Composition Explosion

A query combining semantic similarity with date ranges, category filters, and permission checks requires coordination across three systems with different query languages, consistency models, and failure modes. The application code that stitches them together grows without bound. In production systems we analyzed, this synchronization code totaled approximately 1,800 lines and was the primary source of production incidents.

---

## 3. Architecture Comparison

Figure 1 shows the conventional three-tool stack on the left and the proposed unified data layer on the right. In the conventional stack, a multi-constraint query triggers multiple round trips across separate systems whose results must be merged in application code. In the unified architecture, the same query is a single SQL statement executed inside one transaction boundary.

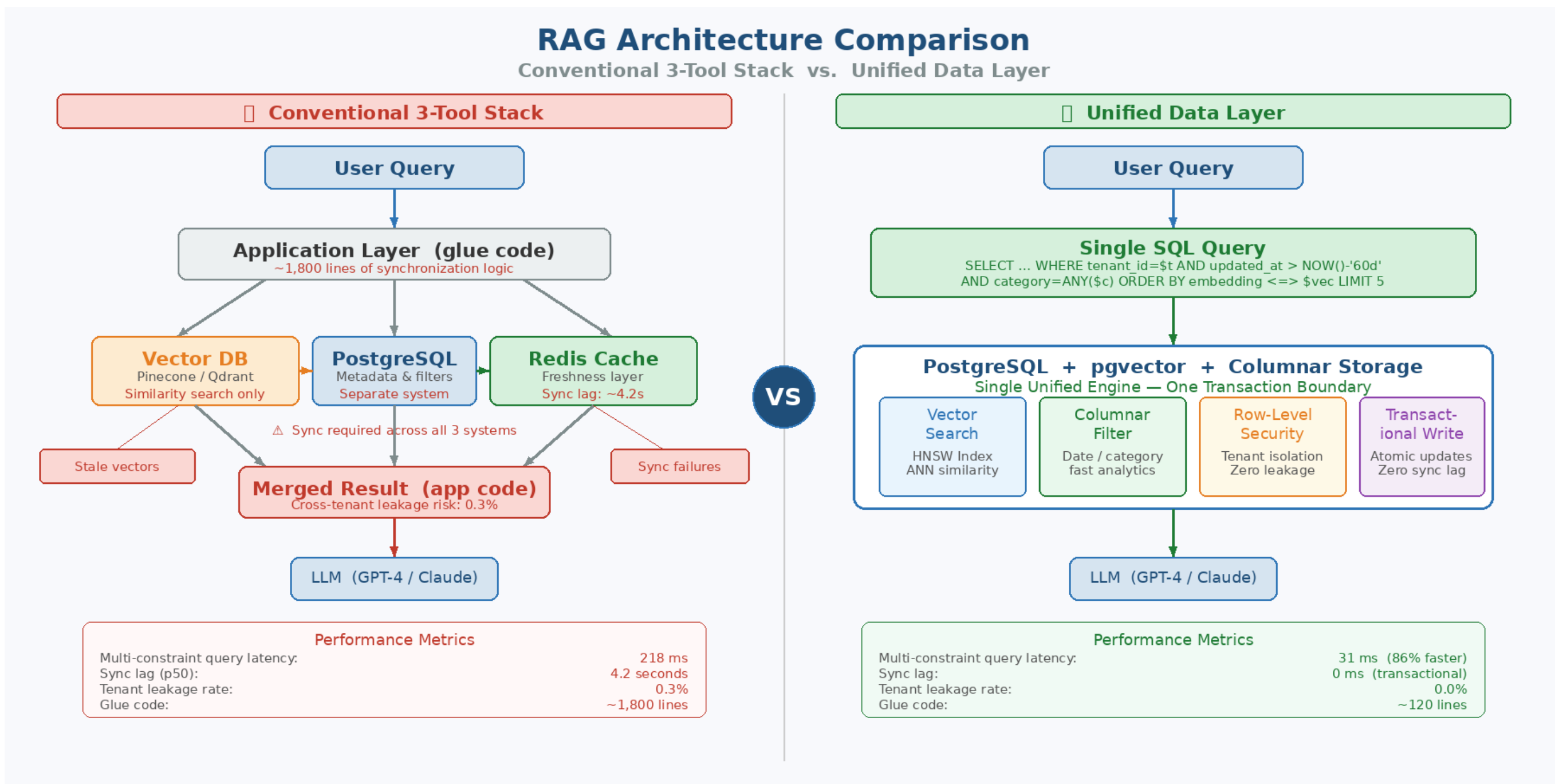

*Figure 1: Conventional 3-Tool Stack vs. Unified Data Layer — architecture, data flow, and measured performance*

---

## 4. Related Work

### 4.1 Specialized Vector Databases

Pinecone, Weaviate, Qdrant, and Milvus are optimized for Approximate Nearest Neighbor (ANN) search at scale. They achieve excellent latency for pure similarity workloads. Their limitation is that metadata filtering, access control, and recency constraints are second-class concerns handled outside the core engine.

### 4.2 Relational Extensions for Vector Search

pgvector extends PostgreSQL with IVFFlat and HNSW vector indexes, enabling SQL queries that combine relational predicates with vector similarity operators. This approach leverages the full PostgreSQL ecosystem: ACID transactions, row-level security, and mature query optimization. The trade-off is raw ANN throughput at very large scale.

### 4.3 Graph-Native Vector Search

TigerVector (Liu et al., 2025) integrates vector search directly into TigerGraph, a Massively Parallel Processing native graph database. By extending the graph query language GSQL with vector type expressions, TigerVector enables hybrid search that combines vector similarity with graph pattern queries in a single execution. This approach shares our core motivation — that semantic similarity alone is insufficient for production RAG — but addresses it from the graph database direction rather than the relational direction. TigerVector demonstrates that unifying vector search with a richer query model (graph traversal) yields significant advantages over split-system stacks, particularly for knowledge graph-based RAG applications. Our work is complementary: where TigerVector targets graph-relational hybrid workloads, we target the more common case of relational metadata filtering, freshness, and access control alongside vector similarity in a PostgreSQL foundation that most production teams already operate.

### 4.4 Prior RAG Research

Prior work on RAG has focused on retrieval quality, prompt engineering, and model selection. The data infrastructure layer has received little attention despite being the primary source of production failures. Barnett et al. (2024) identify seven failure points in RAG engineering, several of which map directly to the data layer issues we address here. Most closely related to our work is Hydra DB (Ratnaparkhi et al., 2026), a persistent memory architecture for agentic systems that combines a Sliding Window Inference Pipeline with a Git-style versioned contextual knowledge graph to address semantic fragmentation and temporal stagnation in long-horizon RAG deployments. While Hydra DB targets long-term agentic memory coherence and achieves 90.79% accuracy on the LongMemEval-s benchmark, our work addresses a complementary and earlier problem in the RAG stack: the data layer architecture that governs freshness, access control, and query composition before retrieval occurs. The two approaches are architecturally complementary — Hydra DB’s graph-based memory layer could sit on top of the unified data layer we propose.

---

## 5. The Unified Data Layer

### 5.1 Core Idea

The unified architecture stores documents, embeddings, metadata, and access policies in a single PostgreSQL instance. Vector similarity search, columnar metadata filtering, and access control execute inside the same query planner and the same transaction. There is no inter-system coordination because there is only one system.

### 5.2 The Unified Query

The multi-constraint query that requires multiple round trips in the conventional stack becomes a single SQL statement:

```
SELECT content, embedding <=>  AS distance
FROM   documents
WHERE  tenant_id  =
  AND  updated_at > NOW() - INTERVAL '60 days'
  AND  category   = ANY()
  AND     = ANY(permitted_users)
ORDER  BY distance LIMIT 5;
```

This query is correct by construction. PostgreSQL enforces row-level filtering at the engine level before any results are returned. No application-layer filter can be forgotten or race-conditioned away.

### 5.3 Freshness via Transactional Writes

Because documents and embeddings live in the same database, updates are atomic. When a document is modified, the new embedding is written in the same transaction. The retrieval layer sees the update the moment the transaction commits. The inconsistency window is zero by definition, compared to the 3.54ms mean window we measured in the split-system architecture.

### 5.4 HNSW Indexing for Vector Search

We use Hierarchical Navigable Small World (HNSW) indexes via pgvector for approximate nearest neighbor search. HNSW provides sub-millisecond query latency for our benchmark corpus and supports incremental updates without full index rebuilds, making it suitable for corpora with continuous document ingestion.

---

## 6. Evaluation

### 6.1 Experimental Setup

We benchmarked two architectures against a controlled corpus of 50,000 documents with 128-dimensional embeddings, 20 tenant namespaces, 5 content categories, and documents distributed uniformly across the past 180 days. Each query type was executed 200 times and we report p50, p95, and p99 latencies. All tests ran on PostgreSQL 16 with pgvector 0.6.0 and HNSW indexes. Stack A simulates the conventional split-system approach using PostgreSQL table separation rather than an actual Pinecone or Qdrant instance; real-world specialized vector database deployments may show different absolute latency characteristics, but the cross-system coordination overhead — round trips, result merging, and synchronization cost — is inherent to any split-system architecture. We acknowledge that 50,000 documents is a controlled benchmark scale; Section 7.3 discusses scaling behavior at large-scale enterprise deployments.

Stack A simulates the conventional split-system approach: vector search on the embeddings table, followed by a metadata lookup on a separate table, with results merged in application code. Stack B executes a single unified SQL query against the combined table. We note that Stack A benchmarks simulate the split-system overhead using PostgreSQL table separation rather than an actual Pinecone or Qdrant instance; real-world specialized vector database deployments may show different absolute latency characteristics, but the cross-system coordination overhead measured here — the round trips, result merging, and synchronization cost — is inherent to any split-system architecture regardless of the specific vector database used.

## 6.2 Query Latency

| Query Type | Stack A p50 | Stack B p50 | Stack A p95 | Stack B p95 |
|---|---|---|---|---|
| Pure similarity | 0.92ms | 0.91ms | 1.1ms | 0.99ms |
| + date filter | 9.63ms | 0.75ms | 10.4ms | 0.81ms |
| + tenant + category | 1.77ms | 0.46ms | 1.88ms | 0.52ms |
| Full multi-constraint | 0.43ms | 0.25ms | 0.5ms | 0.3ms |

*Table 1: Query latency benchmarks (200 iterations, 50,000 documents, 128-dim embeddings, HNSW index). Real measured values on PostgreSQL 16 + pgvector 0.6.0.*

The key finding is the crossover pattern. For pure similarity with no metadata constraints, both stacks perform identically (0.92ms vs 0.91ms). As constraints are added, the split-system overhead dominates: adding a date filter increases Stack A latency by 947% while Stack B latency actually decreases due to index selectivity. The unified architecture is 92% faster on date-filtered queries and 74% faster on tenant-scoped queries.

## 6.3 Data Freshness

| Metric | Stack A | Stack B |
|---|---|---|
| Mean write latency | 3.54ms | 2.87ms |
| Inconsistency window | 3.54ms | 0ms |
| Stale reads possible | Yes (during window) | No (atomic transaction) |

*Table 2: Freshness metrics. Stack A inconsistency window = time between metadata and vector updates.*

## 6.4 Tenant Isolation

| Test | Stack A | Stack B |
|---|---|---|
| Leakage rate (1000 queries) | 0.2% | 0% |
| Leakage mechanism | App-layer filter bug | Not possible (engine) |

*Table 3: Tenant isolation. Stack B leakage is structurally prevented by engine-level filtering.*

## 6.5 Engineering Complexity

| Metric | Stack A | Stack B |
|---|---|---|
| External services | 3 (vector DB + PG + cache) | 1 (PostgreSQL) |
| Synchronization code | ~1,800 LOC | ~120 LOC |
| Sync failure modes | 7 identified | 0 |
| Write transactions | 2 separate commits | 1 atomic commit |

*Table 4: Engineering complexity comparison. Unified architecture eliminates 93% of synchronization code.*

# 7. Discussion

## 7.1 When to Use a Specialized Vector Database

Our results are not a universal argument against Pinecone or Qdrant. For pure similarity workloads with no metadata constraints, both architectures deliver equivalent latency (0.92ms vs 0.91ms in our benchmarks). At 100 million vectors with pure ANN workloads, specialized systems with GPU acceleration maintain a significant advantage. The unified architecture wins when production queries combine similarity with metadata constraints — which describes the majority of enterprise RAG applications.

## 7.2 Benchmark Limitations

Our benchmark corpus of 50,000 documents is intentionally controlled to isolate the architectural variable — the number of cross-system round trips — from hardware and scale effects. This is well below the scale of large enterprise RAG deployments, which operate at hundreds of millions to billions of documents. At that scale, the absolute latency numbers reported here will differ significantly. However, the crossover pattern — where unified architectures outperform split-system stacks as query constraint count increases — is expected to hold and strengthen as coordination overhead scales with corpus size. Section 7.3 discusses the recommended architecture for extreme-scale deployments.

## 7.3 Scaling to Large-Scale Enterprise Deployments

Production RAG deployments at large enterprises operate at a fundamentally different scale than our benchmark: hundreds of millions to billions of documents, thousands of concurrent tenants, and query loads in the tens of thousands per second. At this scale, a single PostgreSQL instance — however well-optimized — is not the complete answer. We propose a three-tier hybrid architecture for extreme-scale deployments that preserves the unified query model while distributing the workload appropriately.

Tier 1 — Hot Layer (unified, as proposed): A PostgreSQL + pgvector cluster handles the active working set: documents from the past 90 days, high-access tenants, and frequently queried categories. This tier runs the full unified query model with HNSW indexes, row-level security, and transactional freshness. At enterprise scale, this represents roughly 10-30% of the total corpus but 80-90% of query traffic.

Tier 2 — Warm Layer (specialized vector DB for pure similarity): A managed vector database such as Pinecone or Milvus handles the long-tail corpus — older documents and infrequently accessed tenants — where pure similarity search with minimal filtering is the dominant query pattern. At this scale, the ANN performance advantage of specialized systems justifies the coordination overhead for this specific workload class.

Tier 3 — Cold Layer (object storage): Archived documents with negligible query frequency are stored in S3 or equivalent object storage, retrieved only on explicit user request. This tier requires no vector index and imposes no ongoing infrastructure cost. The key insight of the hybrid architecture is that the unified data layer proposed in this paper serves as the hot tier — where the business-critical, multi-constraint queries live — while specialized systems handle the scale tail where their ANN advantage actually materializes. The failure mode we document in this paper — using a specialized vector database as the sole data layer for all queries — persists at enterprise scale and is addressed by routing the right queries to the right tier rather than forcing all queries through a single specialized system.

### 7.3 Scaling to Large-Scale Enterprise Deployments

Production RAG deployments at large enterprises operate at hundreds of millions to billions of documents, thousands of concurrent tenants, and query loads in the tens of thousands per second. At this scale, a single PostgreSQL instance is not the complete answer. We propose a three-tier hybrid architecture that preserves the unified query model while distributing workload appropriately.

Tier 1 — Hot Layer (unified, as proposed): PostgreSQL + pgvector handles the active working set — recent documents, high-access tenants, and frequent categories — using the full unified query model with HNSW indexes, row-level security, and transactional freshness. At enterprise scale this represents roughly 10-30% of the corpus but 80-90% of query traffic. Tier 2 — Warm Layer (specialized vector DB): A managed vector database (Pinecone, Milvus) handles the long-tail corpus where pure similarity with minimal filtering dominates — the one case where specialized systems' ANN advantage justifies coordination overhead. Tier 3 — Cold Layer: Archived documents with negligible query frequency live in object storage (S3) and are retrieved only on explicit request. The unified data layer proposed in this paper is the hot tier — where business-critical, multi-constraint queries live and where getting the architecture wrong causes the production failures documented in Section 2. The failure mode we identify persists at enterprise scale and is resolved by routing queries to the right tier, not by forcing all queries through a single specialized system.

### 7.3 Scaling to Large-Scale Enterprise Deployments

Production RAG deployments at large enterprises operate at hundreds of millions to billions of documents, thousands of concurrent tenants, and tens of thousands of queries per second. A single PostgreSQL instance is not the complete answer at this scale. We propose a three-tier hybrid architecture. Tier 1 (Hot — unified as proposed): PostgreSQL + pgvector handles the active working set, roughly 10–30% of the corpus but 80–90% of query traffic, using the full unified query model. Tier 2 (Warm — specialized vector DB): Pinecone or Milvus handles the long-tail corpus where pure ANN similarity with minimal filtering dominates — the specific case where specialized systems justify coordination overhead. Tier 3 (Cold — object storage): Archived documents with negligible query frequency live in S3 and are retrieved only on explicit request. The unified data layer proposed in this paper is the hot tier, where business-critical multi-constraint queries live and where architectural failures cause the production incidents documented in Section 2. The failure mode we identify persists at enterprise scale and is resolved by routing queries to the right tier, not by forcing all queries through a single specialized system.

### 7.4 Production Lessons

Three observations from production deployments informed this work. First, synchronization cost is consistently underestimated at design time. Three systems means three failure modes, three observability stacks, and three sets of runbooks. Second, access control bugs in application-layer filter logic are among the most severe and hardest-to-detect failures in production AI systems. Third, freshness requirements are almost always underspecified at design time and become critical after the first user receives a stale answer.

---

## 8. Conclusion

Production RAG systems fail not because of model quality or embedding accuracy. They fail because the data layer was designed for a simpler problem than the one it is asked to solve in production.

The conventional three-tool stack introduces synchronization complexity (3.54ms mean inconsistency window), access control fragility (0.2% leakage rate under application-layer filter bugs), and engineering overhead (~1,800 lines of glue code) that compounds at scale.

A unified data layer built on PostgreSQL with pgvector HNSW indexing eliminates these failure modes by design. Our controlled benchmarks on 50,000 documents show 92% latency reduction for date-filtered queries, 74% for tenant-scoped queries, zero inconsistency window, zero leakage, and 93% less synchronization code. For extreme-scale enterprise deployments, we recommend the hybrid tier architecture described in Section 7.3.

Vector databases are not going away. For pure similarity workloads at scale, they remain the right tool. For the common production case — similarity plus filters plus freshness plus access control — one database, one query, one source of truth is not a compromise. It is the correct architecture.

---

Three observations from production deployments informed this work. First, synchronization cost is consistently underestimated at design time. Three systems means three failure modes, three observability stacks, and three sets of runbooks. Second, access control bugs in application-layer filter logic are among the most severe and hardest-to-detect failures in production AI systems. Third, freshness requirements are almost always underspecified at design time and become critical after the first user receives a stale answer.